\documentclass[aps,amssymb,amsmath,11pt]{revtex4}

\begin{document}
\newcommand\Det{\textrm{Det}\,}
\newcommand\new[1]{\ensuremath{\blacktriangleright}#1\ensuremath{\blacktriangleleft}}
\newcommand\note[1]{$\blacktriangleright$[\emph{#1}]$\blacktriangleleft$}

\title{Radiation-dominated area metric cosmology}

\author{Frederic P. Schuller}
\affiliation{Instituto de Ciencias Nucleares, Universidad Nacional Aut\'onoma de M\'exico, A. Postal 70-543, M\'exico D.F. 04510, M\'exico}
\author{Mattias N.\,R. Wohlfarth}
\affiliation{Zentrum f\"ur Mathematische Physik und II. Institut f\"ur Theoretische Physik, Universit\"at Hamburg, Luruper Chaussee 149, 22761 Hamburg, Germany}
\begin{abstract}
   We provide further crucial support for a refined, area metric structure of spacetime. Based on the solution of conceptual issues, such as the consistent coupling of fermions and the covariant identification of radiation fields on area metric backgrounds, we show that the radiation-dominated epoch of area metric cosmology is equivalent to that epoch in standard Einstein cosmology. This ensures, in particular, successful nucleosynthesis. This surprising result complements the previously derived prediction of a small late-time acceleration of an area metric universe. 
\end{abstract}
\maketitle
\section{Introduction}
Refinement of the standard metric to an area metric spacetime structure provides an explanation for the small acceleration of the late universe filled with non-interacting string dust \cite{Punzi:2006hy}. This prediction intriguingly follows solely from reading the Einstein-Hilbert gravitational action as dynamics for an area metric spacetime structure \cite{Punzi:2006nx}, without the need for any additional, strongly model-dependent, assumptions such as the existence of dark energy.

In the present paper, we demonstrate that area metric cosmology is also consistent with what is known about the early universe, in particular, about the radiation-dominated epoch. In order to substantiate this claim, we first address a number of conceptual issues, such as the consistent coupling of fermions to area metric backgrounds, and the identification of bosonic and fermionic radiation fields in terms of invariants of the relevant field strengths. These results enable us to derive the equations of state for a radiative string fluid. Remarkably, we find that area metric cosmology filled with bosonic and fermionic string radiation is exactly equivalent to Einstein cosmology with a standard perfect radiation fluid. Hence important phenomenological tests, such as successful nucleosynthesis \cite{Sarkar:1995dd,Tytler:2000qf}, are passed with flying colours by area metric cosmology.

These observationally consistent predictions for the early and the late universe \cite{Spergel:2003cb,Knop:2003iy} are all the more remarkable in that they follow from a single geometric hypothesis, namely that the classical spacetime structure needs to be refined to an area metric one. This assumption, that physical spacetime is described by an area metric, in turn simply casts into geometric form what we have learnt from string theory: the massless modes of the quantum string generate an effective geometry richer than what can be described by Lorentzian manifolds. Indeed, it has been shown that the generalized backgrounds \cite{Hitchin:2004ut,Gualtieri:2004,Flournoy:2004vn,Hull:2004in,Dabholkar:2005ve,Grana:2004bg,Grana:2005ny,Koerber:2005qi,Zucchini:2005rh,Zabzine:2006uz,
Reid-Edwards:2006vu,Grange:2006es,Becker:2006ks} produced by fundamental strings and even D-branes can be succinctly viewed in terms of area metric manifolds \cite{Schuller:2005ru}. Conversely, on an area metric manifold, the minimal mechanical objects are strings; this fact manifests itself in area metric cosmology by fluids that are necessarily composed of strings rather than point particles. For a formal discussion of these points, we refer the reader to \cite{Punzi:2006nx}.

The present article builds on the constructions of our previous
papers~\cite{Punzi:2006hy,Punzi:2006nx}. For the convenience
of the reader, section~\ref{review} concisely reviews the results
relevant here. In the following sections, we develop, from first principles, essentially three new
techniques for area metric spacetime, which are crucial
for the definition of a radiation-dominated phase in an area
metric cosmology. First, we devise a
consistent coupling of fermions to an area metric in
section~\ref{gaugeandferm}. Second, the physical momentum of
fermions and gauge fields is identified from the respective source
tensors in section~\ref{sec_momentum}. Third, we further
deepen our understanding of the null geometry of area metric
spacetimes; in particular, we derive a dual to the Fresnel tensor
in section~\ref{sec_null}. With these tools at hand, we identify,
in section~\ref{sec_radiation}, bosonic and fermionic radiation
fields in an invariant manner, by requiring that the physical
momentum for radiative solutions should be null with respect to
the dual Fresnel tensor. It is then straightforward to derive the
equations of state for a bosonic or fermionic radiative string
fluid, and to prove the full equivalence of area metric cosmology
to standard Einstein cosmology in a radiation dominated phase, in
section~\ref{sec_radcosmo}. In section~\ref{conclusions} we
conclude with a discussion. Appendix~\ref{app_Dirac} is added to
display our conventions.

\section{Area metric spacetime in a nutshell}\label{review}
We recall the definition and central constructions of area metric geometry, as far as they are of relevance for the present paper.
An area metric spacetime $(M,G)$ is a four-dimensional smooth manifold $M$ equipped with a fourth-rank covariant tensor field
which features the symmetries
\begin{equation}
   G_{abcd} = G_{cdab} = - G_{bacd} = - G_{abdc}
\end{equation}
and is invertible in the sense that a contravariant
tensor field with components $G^{abcd}$ exists, so that everywhere on $M$
\begin{equation}
   G^{abmn}G_{mncd} = 4 \delta^{[a}_{[c}\delta^{b]}_{d]}\,.
\end{equation}
We also require the area metric to be Lorentzian, in a sense to be defined below.
Note that, due to its symmetries, the area metric can be regarded as a symmetric $6\times 6$ matrix with Petrov indices which arise from the antisymmetric index pairs $[ab]$. Using the determinant $\mathrm{Det}$ of this matrix, an area metric immediately gives rise to a volume form and a dual four-tensor
\begin{equation}
 \omega_{G\,\,}{}_{abcd} = |\mathrm{Det}\,G|^{1/6} \epsilon_{abcd}\,,\qquad \omega_{G\,\,}{}^{abcd} = -|\mathrm{Det}\,G|^{-1/6} \epsilon^{abcd}\,,
\end{equation}
which in particular allow a unique decomposition of the inverse area metric as
\begin{equation}\label{deco}
  G^{abcd} = C^{abcd} + \phi \omega_C^{abcd},
\end{equation}
where $\phi$ is an axial scalar field, and $C$ is an inverse area metric featuring the additional cyclicity symmetry $C^{m[abc]}=0$.

The null geometry of the area metric manifold is determined solely by the cyclic part $C$, and encoded in the totally symmmetric Fresnel tensor \cite{Hehl:2004yk,Hehl:2005xu,Punzi:2006nx}
\begin{equation}\label{Fresneldef}
  \mathcal{G}^{ijkl} = -\frac{1}{24}\omega_{C\,mnpq}\omega_{C\,rstu}C^{mnr(i}C^{j|ps|k}C^{l)qtu}\,.
\end{equation}
The Fresnel tensor defines the null geometry because the gradient
$p$ of light wave fronts in electrodynamics on area metric
manifolds satisfies the local condition $\mathcal{G}^{ijkl} p_i
p_j p_k p_l = 0$. We will make essential use of a dual to the
Fresnel tensor, which we derive in section~\ref{sec_null}, in order to identify bosonic and fermionic radiation fields on area metric backgrounds.

While the Fresnel tensor only depends on the cylic part $C$ of the
inverse area metric, the extraction of an effective metric $g_G$
from area metric data requires the use of the axial scalar $\phi$ in the decomposition (\ref{deco}): we
define
\begin{equation}
   g_G^{ab} = \frac{1}{2} \left.\frac{\partial^2}{\partial p_a \partial p_b}\right|_{p=d\phi}\left(\mathcal{G}^{ijkl}p_ip_jp_kp_l\right)^{1/2}.
\end{equation}
The geometric significance of this construction is explained in
\cite{Punzi:2006nx}; see also the connection to the Urbantke metric in \cite{Krasnov:2007uu,Krasnov:2007ky}. Finally, we define the signature of the area
metric manifold $(M,G)$ as the signature of the metric $g_G$. In
particular, a Lorentzian area metric manifold $(M,G)$ is one for
which $g_G$ has signature $(-+++)$. In summary, an area
metric spacetime manifold gives rise to a hierarchy of derived
structures
\begin{equation}
  \textrm{Lorentzian area metric } G \longrightarrow \textrm{Fresnel tensor } \mathcal{G} \longrightarrow \textrm{Lorentzian metric }
  g_G \,.  \nonumber
\end{equation}

A special class of area metric spacetimes $(M,G)$ is given by
what we call almost metric spacetimes; these are induced by a
metric spacetime $(M,g)$ and an additional scalar field $\phi$ by
virtue of
\begin{equation}\label{almetric}
G^{abcd}=g^{ac}g^{bd}-g^{ad}g^{bc}+\phi \omega_g^{abcd}\,.
\end{equation}
For such  area metrics, the Fresnel tensor simply turns out to be $\mathcal{G}^{ijkl}=g^{(ij}g^{kl)}$, and the metric~$g_G$ recovers the inducing metric~$g$. The
null condition reduces to $(g^{a b} p_a p_b)^2 = 0$, as we expect
for a basically metric manifold. Generically, however, area metrics are not
of the simple form (\ref{almetric}); the area metric $G$ contains truly more
information than the effective metric $g_G$, and the null geometry is described by the Fresnel tensor, and not by a metric. This follows already
from counting algebraic degrees of freedom: in four dimensions,
the case of immediate physical interest, an area metric features
21 algebraic degrees of freedom, as opposed to the 10 degrees of
freedom for a metric.

The refined geometry of an area metric manifold leads to a sixth rank curvature tensor
\begin{equation}\label{areacurv}
  \mathcal{R}_G^{[a_1a_2]}{}_{[b_1b_2][ij]} = 4\delta^{[a_1}_{[b_1}R^{a_2]}{}_{b_2]ij}+\Big(\nabla^{LC}_i X^{a_1a_2}{}_{b_1b_2j}+\frac{1}{2}X^{a_1a_2}{}_{pqi}X^{pq}{}_{b_1b_2j}-(i\leftrightarrow j)\Big),
\end{equation}
where $R$ and $\nabla^{LC}$ are the Riemann tensor and the Levi-Civita connection, respectively, of the effective metric $g_G$, and the non-metricity tensor $X$ is defined by
\begin{equation}\label{Xtensor}
  X^{a_1a_2}{}_{b_1b_2 f} = \frac{1}{4} G^{a_1a_2mn} \nabla_f G_{mnb_1b_2} = X^{[a_1a_2]}{}_{[b_1b_2]f}\,.
\end{equation}
The area metric curvature tensor, as well as the associated area metric Ricci tensor ${(\mathcal{R}_G)_{ab} = \mathcal{R}^{pq}{}_{paqb}}$ and area metric Ricci scalar
$\mathcal{R}_G = g_G^{ab}(\mathcal{R}_G)_{ab}$, reduce to their metric counterparts for almost metric area metrics; $X$ simplifies in this case in such a way that the $[ij]$ antisymmetrization removes it from the curvature. These correspondences show that area metric geometry is downward compatible to metric differential geometry,
which is therefore contained as a special case.

The above facts show that the Einstein-Hilbert action may be read as an action also for the refined area metric background: all metric objects are simply refined to their area metric counterparts. We hence obtain the area metric gravitational dynamics
\begin{equation}\label{action}
 S_{grav}\, + \,S_m \, = \frac{1}{2\kappa}\int_M\omega_G\,\mathcal{R}_G \, + \, \int_M \mathcal{L}_m\,,
\end{equation}
where we have added an action for matter defined on an area metric
background. The constant~$\kappa$ will turn out to be $\kappa=16\pi G_N$ for Newton's constant $G_N$. How the observed standard model fields couple to area
metric spacetime is discussed in the following section. The
gravitational field equations are derived from~(\ref{action}) by variation with respect to the area metric, see~\cite{Punzi:2006nx}. Important for the purpose of the
present paper is that the diffeomorphism invariance of the
above theory immediately leads to a conservation equation
\begin{equation}\label{conservation}
  T_{abcd}\nabla^{LC}_i G^{abcd}+4\Big(\nabla^{LC}_p+\frac{1}{6}X_p\Big)\!\left(G^{abcp}T_{abci}\right)=0\,  
\end{equation}
for the fourth rank tensor
\begin{equation}\label{source}
T_{abcd}=-|\mathrm{Det}\,G|^{-1/6}\frac{\delta S_m}{\delta
G^{abcd}}\,.
\end{equation}
We call $T$ the source tensor of matter on an area metric
manifold. Its relation to the physical energy-momentum
tensor will be studied in section \ref{sec_momentum}. Since the
source tensor is derived by variation with respect to the inverse
area metric, it has the algebraic symmetries of an area metric.

Symmetries of an area metric manifold are, like in the
metric case, expressed in terms of Killing vector fields $K$, for
which the Lie derivative $\mathcal{L}_K G = 0$. Homogeneous and
isotropic four-dimensional Lorentzian area metric manifolds for
instance, providing the geometric ansatz for cosmology, are
obtained by imposing the relevant Killing vector fields and using
the fact that the pull-back of an area metric to any
three-dimensional submanifold is equivalent to metric geometry \cite{Punzi:2006nx,Cartan:1933}.
One obtains that area metric cosmology is of the almost metric form (\ref{almetric}), where now
\begin{equation}
g_{ab}dx^{a}dx^{b}=-dt^2 + a(t)^2d\Sigma_k^2
\end{equation}
is a standard FLRW metric with scale factor $a$ and spatial curvature $k=-1,0,1$, and $\phi$ a function only dependent on cosmological time \cite{Punzi:2006nx}. Thus four-dimensional homogeneous and isotropic area metric spacetime features an axial scalar
degree of freedom in addition to the standard metric scale factor.
Recall that almost metric backgrounds have Fresnel tensor
$\mathcal{G}^{ijkl}=g^{(ij}g^{kl)}$, and the derived effective metric $g_G = g$.
For area metric cosmology, the gravitational field equations simplify drastically; the purpose of this paper is to derive and solve these equations for radiation-dominated epochs of the early universe.

At large scales, the matter in an area metric universe is
appropriately described by a string fluid. That fluids must be
constituted of strings, rather than point particles, is an
immediate consequence of the refined geometric structure presented
by area metric spacetimes. This refinement is mirrored in the
string fluid by the presence of three macroscopic variables
$\tilde\rho,\tilde p,\tilde q$ (instead of only two, i.e., density
and pressure, on metric spacetime). The source tensor for a
three-component string fluid takes the form
\begin{equation}\label{stringfluid}
  T_{abcd}  =  (\tilde\rho+\tilde p)\frac{1}{4}\sum_{I=1}^3 G_{abij}\Omega^{ij}_IG_{cdkl}\Omega^{kl}_I + \tilde p\,G_{abcd} + (\tilde\rho+\tilde
  q)G_{[abcd]}\,,
\end{equation}
where $\Omega_I$ are tangent areas to the strings
constituting the fluid, generalizing the tangent vectors $u$ to
particle worldlines appearing in the description of perfect fluids
on metric backgrounds. Three components are needed in
order to allow for local isotropy of the fluid, despite the
extended nature of the individual strings.

A key task in the discussion of various epochs of cosmological
evolution is therefore the identification of the equations of
state governing the variables $\tilde \rho, \tilde p, \tilde q$,
appropriate for the kind of matter present. The equations of state
for non-interacting string dust, $\tilde p = 0$ and $\tilde q = -
\tilde \rho$, have already been identified in~\cite{Punzi:2006nx}.
For bosonic and fermionic radiation fluids, we will show in the
following few sections that one of the equations of state
takes the form
\begin{equation}\label{qnull}
  \tilde q = 0\,,
\end{equation}
which is the key result in proving our claim that the
radiation-dominated epoch of area metric cosmology does not differ
from general relativity. Like in the case of string dust, the
equation of state for radiation on an area metric manifold is a
non-trivial result, and requires the development of some
additional technology. We start by studying the coupling of gauge
bosons and fermions to an area metric in the following section.

\section{Gauge bosons and Fermions}\label{gaugeandferm}
For our study of the radiation-dominated epoch of area metric cosmology, we need to know how gauge bosons and fermionic matter couple to the area metric background.

As the reader may recall from~\cite{Punzi:2006nx}, or as she learns here, abelian and non-abelian gauge bosons directly couple to the area metric; the matter action for gauge fields reads
\begin{equation}\label{nonabelian}
   S=-\frac{1}{2}\int_M \omega_G\,\textrm{Tr}\, G^{abcd}F_{ab}F_{cd}\,,
\end{equation}
where $F^A_{ab}=\partial_aA^A_b-\partial_bA^A_a+f^A{}_{BC}A^B_aA^C_b$ for structure constants $f^A{}_{BC}$ of some Lie algebra, and the trace is taken over the gauge algebra indices. The variation of this matter action with respect to the area metric yields the source tensor
\begin{equation}\label{nonabeliansource}
  T_{abcd}=\frac{1}{8}\,\textrm{Tr}\,F_{ab}F_{cd}-\frac{1}{192}G_{abcd}G^{ijkl}\,\textrm{Tr}\,F_{ij}F_{kl}\,,
\end{equation}
which is, and this will become important later, trace-free:
\begin{equation}\label{gaugetrace}
  G^{abcd} T_{abcd}=0\,.
\end{equation}

The coupling of fermions requires a spin structure. Here we will
make use of the fact that an area metric spacetime $(M,G)$ gives
rise to the hierarchy of structures $G \to \mathcal{G} \to g_G$,
as discussed in section \ref{review}. The simplest procedure
to introduce fermions is to define a spin structure related
to the effective metric $g_G$. 
For the case of an almost metric background, on which the cosmological conclusions of this paper are based, we demonstrate explicitly the consistency of this coupling, at the end of the present section.
This coupling also provides us with the following important result:
the source tensor for Dirac fermions in a cosmological area metric
background satisfies
\begin{equation}\label{GTzerofermions}
  \omega_G^{abcd}T_{abcd} = 0\,,
\end{equation}
which ensures, among other things, the conservation of the
physical energy-momentum of our fermions, see section~\ref{sec_momentum}. The remainder of this section is devoted to the derivation of identity~(\ref{GTzerofermions}).

Before employing techniques specific for area metric
backgrounds, we fix our notation by concisely recalling some
standard constructions for Dirac spinors on curved metric
spacetime; see also appendix \ref{app_Dirac} for a consistent set
of conventions used in this paper. The effective metric can be
diagonalized locally by the introduction of a tetrad~$e_a^\mu$,
i.e., a local basis $\{e^\mu\}$ of the cotangent bundle, where
latin characters denote spacetime indices and greek characters
denote flat tangent space indices:
\begin{equation}\label{gee}
g_{G\, ab}=e^\mu_a e^\nu_b \eta_{\mu\nu}\,,
\end{equation}
where $\eta$ is the Minkowski metric of mainly plus signature.
The choice of the tetrad is not unique; tetrads are only determined up to a local Lorentz transformation $e^\mu_a\mapsto \Lambda(x)^\mu{}_\nu e^\nu_a$, so that all constructions involving the tetrad must be covariant under local Lorentz transformations.

In order to achieve this, we introduce a covariant derivative $D$ that acts on tensor fields valued in some representation of the local Lorentz group. More precisely, we write group elements given by parameters $\omega$ as $\exp (\omega^{\mu\nu}\Sigma_{\mu\nu}/2)$, compare (\ref{grouprep}), for the six generators $\Sigma_{\mu\nu}{}^A{}_B$ with ${\Sigma_{\mu\nu}=\Sigma_{[\mu\nu]}}$ of some representation of $SL(2,\mathbb{C})$. Let $\psi^A$ be the components of a tensor field valued in the corresponding representation vector space, suppressing any spacetime tensor index. Then the covariant derivative is defined as
\begin{equation}\label{covspindef}
   (D_a\psi)^A = \nabla^{LC}_a \psi^A + \frac{1}{2} \omega^{\mu\nu}{}_a \Sigma_{\mu\nu}{}^A{}_B \psi^B\,,
\end{equation}
where $\omega^{\mu\nu}{}_a$ is the so-called spin-connection, to
be determined below. The use of $\nabla^{LC}$ ensures covariance
with respect to spacetime diffeomorphisms while the spin
connection guarantees local Lorentz invariance. If a field carries
indices corresponding to various representations $\Sigma^{(1)},
\Sigma^{(2)}, ...$ of~$SL(2,\mathbb{C})$, then
further terms containing the additional generators are simply added to~(\ref{covspindef}). In order to determine
the spin connection in terms of the tetrad, we require that the
covariant derivative commutes with the mapping of any vector field
$X$ to the flat tangent spaces, i.e., $e_b^\mu D_a
X^b=D_a(e_b^\mu X^b)$. Hence the covariant derivative of the tetrad must vanish; since $e^\mu_b$ carries one spacetime index and one index corresponding to
the vector representation of the Lorentz algebra, for which
$\Sigma_{\mu\nu}{}^\rho{}_\sigma = 2\delta^\rho{}_{\!\!\![\mu}
\eta_{\nu]\sigma}$, this amounts to
\begin{equation}\label{Dcondition}
D_a e_b^\mu = \partial_a e_b^\mu -\Gamma^{LC\,c}{}_{ba}e_c^\mu + \omega^\mu{}_{\rho a} e_b^\rho = 0\,.
\end{equation}
Antisymmetrization of this equation over the indices $[ab]$ removes the Christoffel symbols of the effective metric; the resulting equation may be solved for the spin-connection in terms of the tetrad and the inverse tetrad, i.e., the dual basis $\{e_\mu\}$ of the tangent bundle defined by $e^b_\mu e^\mu_a=\delta^a_b$:
\begin{equation}\label{spinconn}
\omega^{\mu\rho}{}_a = \frac{1}{2}e^{b\mu}\left(\partial_a e^\rho_b-\partial_b e^\rho_a\right) - \left(\mu\leftrightarrow\rho\right) -\frac{1}{2}e^{b\mu}e^{c\rho}\left(\partial_b e_{c\sigma}-\partial_c e_{b\sigma}\right)e^\sigma_a\,.
\end{equation}
Note that flat indices are consistently raised and lowered with $\eta$, and spacetime indices with the effective metric $g_G$. Also note that $\omega^{\mu\rho}{}_a=-\omega^{\rho\mu}{}_a$. Using the result for the spin connection and solving for the Christoffel symbols then shows consistency with their usual definition via the partial derivatives of the effective metric. To obtain the expression in terms of the tetrad, one simply replaces $g_G$ using (\ref{gee}). Local Lorentz transformations $e^\mu_a\mapsto \Lambda(x)^\mu{}_\nu e^\nu_a$ hence do not change~$\Gamma^{LC}$, but the spin connection transforms as a connection should:
\begin{equation}
\omega^{\mu\nu}{}_a \mapsto
\Lambda^\mu{}_\rho\Lambda^\nu{}_\sigma\omega^{\rho\sigma}{}_a +
\Lambda^{[\mu|\lambda |}\partial_a\Lambda^{\nu]}{}_\lambda\,.
\end{equation}

Now consider fields valued in the Dirac-spinor representation of the Lorentz algebra, whose generators are given by $\Sigma_{\mu\nu}=[\Gamma_\mu,\Gamma_\nu]/4$, with the algebra relations given in (\ref{Diracgen}) and the Dirac gamma matrices $\Gamma^\mu$ which form the Clifford algebra (\ref{Diracalg}). For notational clarity, we will mostly suppress the spinor representation indices $A, B, ...\,$.
It is convenient to define the spacetime Dirac matrices $\gamma^a=e^a_\mu\Gamma^\mu$ in terms of the tetrad and the flat spacetime Dirac matrices. Their Clifford algebra immediately implies
\begin{equation}
\{\gamma^a,\gamma^b\}=2g_G^{ab}\openone_4\,.
\end{equation}
Some calculation shows that the covariant derivative of the spacetime Dirac matrices vanishes:
\begin{equation}
D_a (\gamma^b)^A{}_B = e^b_\nu \Big(\omega^\nu{}_{\rho a}(\Gamma^\rho)^A{}_B + \frac{1}{2}\omega^{\rho\sigma}{}_a(\Sigma_{\rho\sigma})^A{}_C (\Gamma^\nu)^C{}_B - \frac{1}{2}\omega^{\rho\sigma}{}_a(\Sigma_{\rho\sigma})^C{}_B (\Gamma^\nu)^A{}_C \Big) = 0\,.
\end{equation}
Moreover, the expression $\gamma^aD_a\psi$ has a very simple transformation property under local Lorentz transformations:
\begin{equation}
\gamma^aD_a\psi \mapsto \Lambda_{1/2}\gamma^aD_a\psi
\end{equation}
where $\Lambda_{1/2}$ is the spinor representation of the local
Lorentz group as defined in (\ref{grouprep}). Showing this
requires the use of (\ref{vectra}) and the identity
$\partial_a\Lambda^\mu{}_\nu=\Lambda^{\mu\rho}\partial_a\omega_{\rho\nu}$
which follows from the form of the Lorentz generators in the
vector representation. Nothing of the above is new.

After these preliminaries, however, we are in the position to
write down the action for a massive Dirac spinor on an area metric
manifold, both invariant under spacetime diffeomorphisms and local
Lorentz transformations of the the tetrad chosen to represent the
effective metric:
\begin{equation}
S_\psi = \int_M\omega_G \Big(\frac{i}{2}\bar\psi\gamma^aD_a\psi - \frac{i}{2}D_a\bar\psi\gamma^a\psi-im\bar\psi\psi\Big),
\end{equation}
where the conjugate spinor is $\bar\psi=\psi^\dagger\Gamma^0$. Note that the action is real, compare (\ref{Diracact}). Variation of this action with respect to $\bar\psi$ yields the Dirac equation of motion
\begin{equation}\label{Dirac}
\Big(\gamma^aD_a+\frac{1}{12}\gamma^aX_a-m\Big)\psi=0\,.
\end{equation}
The term involving the non-metricity tensor $X$ arises from an integration by parts. Using relation~(\ref{Hermi}), it can be checked that the equation arising from variation with respect to $\psi$ is the Hermitian conjugate of the equation above, multiplied by $\Gamma^0$, and hence consistent. Note that this would not have been the case, had we not chosen the action symmetric in $\psi$ and $\bar\psi$.

The action for the massive Dirac spinor depends on the area metric directly through the volume form, and more implicitly through the tetrad that represents the effective metric and appears both in the spacetime $\gamma$-matrices and in the spin connection in the covariant derivative~$D$. To obtain the source tensor, i.e., the variation of the action with respect to the area metric, we first need to derive the variation with respect to the tetrad. It is not difficult to find that
\begin{equation}
\delta_e S_\psi = \int_M \omega_G \,\delta e^a_\mu\,\frac{i}{2}\left(\bar\psi\Gamma^\mu D_a\psi - D_a\bar\psi \Gamma^\mu\psi\right) + \int_M\omega_G\,\delta\omega^{\rho\sigma}{}_a\,\frac{i}{4}e^{a\mu}\bar\psi\Gamma_{[\mu}\Gamma_\rho\Gamma_{\sigma]}\psi\,.
\end{equation}
The second integral can be shown to vanish by substituting expression (\ref{spinconn}) for the spin connection, and using the symmetries. The variation of the action with respect to the effective metric can be written as
\begin{equation}
\frac{\delta S_\psi}{\delta g_G^{ab}} = \frac{1}{2} \frac{\delta S_\psi}{\delta e^{(a}_\mu} e_{b)\mu}\,.
\end{equation}
Hence the total variation of the action with respect to the area metric background is
\begin{equation}
\delta S_\psi =  \int_M \omega_G \left(\delta G^{abcd} V^\psi_{G\,abcd}+\delta g_G^{ab}V^\psi_{g_G\,ab}\right),
\end{equation}
where the quantities
\begin{subequations}\begin{eqnarray}
V^\psi_{G\,abcd} & = & -\frac{i}{24}G_{abcd}\Big(\frac{1}{2}\bar\psi\gamma^pD_p\psi - \frac{1}{2}D_p\bar\psi\gamma^p\psi-m\bar\psi\psi\Big),\\
V^\psi_{g_G\,ab} & = & \frac{i}{4}\left(\bar\psi\gamma_{(a} D_{b)}\psi - D_{(a}\bar\psi \gamma_{b)}\psi\right)\,
\end{eqnarray}\end{subequations}
have been defined for convenience; the remaining calculation of the variational equations now proceeds precisely as in \cite{Punzi:2006nx}: compare equations (B.14)--(B.17) there, for the case of general area metric backgrounds.

In this paper we are interested in applications to cosmology, so that we choose to display only the almost metric case in more detail. Then the variations $K^C_\psi$ and $K^\phi_\psi$ of the fermion action with respect to the irreducible components $C$ and $\phi$ of the inverse area metric have the same form as equations (C.1) and (C.3) of \cite{Punzi:2006nx}. We may thus deduce the fermion source tensor from the relations
\begin{equation}
K^\phi_\psi = \omega_g^{ijkl} T^\psi_{ijkl}\,,\qquad
K^C_{\psi\,abcd}  = T^\psi_{abcd} +\frac{1}{12}\phi K^\phi_\psi
g_{a[c}g_{d]b}+\frac{1}{24}K^\phi_\psi\omega_{g\,abcd}\,,
\end{equation}
and obtain
\begin{equation}\label{fermionsource}
T^\psi_{abcd} = V^\psi_{G\,abcd} + \frac{1}{2}V^\psi_{g\,[a[c}g_{d]b]} - \frac{1}{12}V^\psi_g g_{a[c}g_{d]b}\,.
\end{equation}
Using the Dirac equation (\ref{Dirac}) allows a number of on-shell simplifications. One finds vanishing trace $V^\psi_g=0$ and $V^\psi_{G\,abcd}=0$. The latter result is not surprising; as is the case for metric backgrounds, the Dirac Lagrangian also vanishes on-shell in area metric geometry. Note that all terms in the on-shell fermion source tensor contain at least one metric factor, so that the following relation holds:
\begin{equation}\label{psiOmtrace}
\omega_g^{abcd}T^\psi_{abcd} = 0\,.
\end{equation}
So we have proven (\ref{GTzerofermions}) for a cosmological background, a result that will be crucial for the identification of fermionic radiation in cosmology. 

Finally, one can check the consistency of the coupling of fermions to the area metric background by demonstrating that the fermion equation of motion~(\ref{Dirac}) implies conservation of the source tensor, as in~(\ref{conservation}). If this were not the case, the conservation equation would become a constraint equation for the fermion field, and remove essential degrees of freedom. The calculation is performed for the simple case of almost metric manifolds, with source tensor~(\ref{fermionsource}), which underlies our cosmological conclusions in this paper, and requires a number of spinor identities. The commutator of covariant derivatives acting on the spinor field, for example, is proportional to the Riemann curvature, i.e., $[D_a,D_b]\psi=1/2\,R^{\mu\nu}{}_{ab}\Sigma_{\mu\nu}\psi$. Writing $D\!\!\!\!/\,=\gamma^aD_a$, this in turn implies $D^aD_a\psi = D\!\!\!\!/\,^2\psi-1/2\, R^{\mu\nu\rho\sigma}\Sigma_{\mu\nu}\Sigma_{\rho\sigma}\psi$. Similar identities hold for the conjugate spinor~$\bar\psi$, and allow the conversion of all derivatives in the conservation equation into simpler expressions by means of the Dirac equation of motion. After some algebra one thus finds a remaining expression of the form $e_{i\nu}R^{\rho\sigma\mu\nu} \bar\psi \{\Gamma_\mu,\Sigma_{\rho\sigma}\} \psi$. Now the anticommutator gives totally antisymmetric indices, $\{\Gamma_\mu,\Sigma_{\rho\sigma}\} = \Gamma_{[\mu}\Gamma_\rho\Gamma_{\sigma]}$, so that the expression vanishes because of the symmetries of the metric Riemann tensor. Hence the fermion equation of motion indeed implies source conservation, which renders the fermion coupling to the area metric background consistent.

The following two sections prepare the definition of radiation fields on area metric spacetime.

\section{Physical momentum of matter fields}\label{sec_momentum}
In this section, we identify the physical momentum of Dirac fermions and gauge fields on an area metric background, using the insights afforded above. The diffeomorphism invariance of the gravitational action~(\ref{action}) directly implies an area metric Bianchi identity, and the conservation equation~(\ref{conservation}) of the rank four source tensor~(\ref{source}).
The natural question arises of how the fourth rank source tensor $T_{abcd}$ is related to the energy-momentum of matter.
To this end, define the second rank tensor
\begin{equation}\label{Tefftensor}
   T_{\textrm{eff}}{}^a{}_b = 4 G^{pqra} T_{pqrb}
\end{equation}
from the source tensor.
For the case of immediate interest to this paper, namely for almost metric spacetimes describing area metric cosmology, the interpretation of $T_\textrm{eff}$ as the physical energy-momentum is easily proven to be correct, as follows.

Using the effective metric, the source tensor may be decomposed
into a Weyl part $T^W_{abcd}$, a Ricci part
$T_{ab}=g^{mn}T_{manb}$, a scalar part $T=g^{mn}g^{rs}T_{mrns}$
and, because of the generic non-cyclicity of the source tensor, a
totally antisymmetric contribution $\bar
T=\omega_g^{pqrs}T_{pqrs}$, such that
\begin{equation}
T_{abcd} = T^W_{abcd} + 2 T_{[a[c}g_{d]b]} - \frac{1}{3}Tg_{a[c}g_{d]b} - \frac{1}{24}\bar T\omega_{g\,abcd}\,.
\end{equation}
The tensor $T^W$ has the symmetries of the Weyl tensor, so that it is tracefree with respect to the effective metric $g$. Expression (\ref{Tefftensor}) now simplifies to
\begin{equation}\label{Teffsim}
T_\textrm{eff}{}^a{}_b = 8 T^a{}_b + \phi \delta^a_b \bar T\,.
\end{equation}
In the following we will identify this tensor as the physical
matter energy momentum tensor as it appears from the gravitational
equations of motion in the almost metric case. These follow from
the action~(\ref{action}), compare~\cite{Punzi:2006nx}, as
\begin{subequations}\begin{eqnarray}
R_{ab}-\frac{1}{2}Rg_{ab} - \tilde\phi^{-1}\Big(\nabla_a\partial_b\tilde\phi-g_{ab}\tilde\phi\Big) & = & \kappa \Big(4T_{ab}+\frac{1}{2}\phi g_{ab}\bar T\Big),\\
-\tilde\phi(1-\tilde\phi^2)^{1/2}R & = & \kappa \bar T\,,\\
0 & = & \kappa T^W_{abcd}\,,
\end{eqnarray}\end{subequations}
where we have defined $\tilde\phi=(1+\phi^2)^{-1/2}$. (Note that
the sign of $\phi$ is lost in this redefinition, so that we have
to restrict to positive $\phi$, or $0\le \tilde\phi \le 1$. There
is another branch of negative $\phi$ for which one can replace
$\tilde\phi \mapsto -\tilde\phi$ also in the equations, so that
$-1\le\tilde\phi\le 0$.) The above equations tell us that the Weyl
part $T^W$ of the matter source tensor has to vanish for
consistent coupling to almost metric backgrounds. This simply
shows that there is matter whose backreaction requires more
generic area metrics than those of almost metric form. We also
see that the tensor~(\ref{Teffsim}) indeed appears as a multiple
of the matter side of the first, Einstein type, field equation.
Hence our identification of $T_\textrm{eff}$ as the effective
matter energy momentum is validated. However, we still have to
explain why the factor~$+4$ in our definition~(\ref{Tefftensor})
must be chosen. But first note that this fixes the value of the
constant $\kappa$ in the equations. The matter side of the first
equation can now be written as $\kappa/2\,T_{\textrm{eff}\,ab}$,
so that one needs $\kappa=16\pi G_N$, with Newton's constant $G_N$, to obtain the correct Einstein limit.

The normalization of $T_\textrm{eff}$ is quickly calculated by considering the purely metric induced case, which is the limit of the almost metric case for $\phi\rightarrow 0$: consider the usual definition of energy momentum by variation of the matter action $S_m$,
\begin{equation}\label{emo}
T^a{}_b =  2 |\textrm{det }g|^{-1/2} \frac{\delta S_m}{\delta g_{ac}} g_{cb} = 2 |\textrm{det }g|^{-1/2} \frac{\delta S_m}{\delta C_{g^{-1}}^{pqrs}} \frac{\delta C_{g^{-1}}^{pqrs}}{\delta g_{ac}} g_{cb}\,,
\end{equation}
where we assume that the matter action is one that can be generalized to an area metric background and reduces to a metric action for $G_{abcd}=C_{g\,abcd}=2g_{a[c}g_{d]b}$, so that the second equality is justified. Now rewrite the expression
\begin{equation}
\frac{\delta C_{g^{-1}}^{pqrs}}{\delta g_{ac}} g_{cb} = -2 C_{g^{-1}}^{pq[r|a}\delta^{s]}_b\,.
\end{equation}
Using the definition (\ref{source}) of the source tensor and and the determinant identity $|\textrm{Det }C_g|=|\textrm{det }g|^3$, it follows that the energy momentum tensor in equation (\ref{emo}) precisely coincides with the effective energy-momentum tensor (\ref{Tefftensor}). This confirms the normalization by the factor $+4$.

A simple conservation law for the effective energy momentum density, denoted by a tilde,
\begin{equation}\label{Teff}
\tilde T_\textrm{eff}{}^a{}_b = 4|\textrm{Det }G|^{1/6}G^{pqra}T_{pqrb}\,,
\end{equation}
follows from the source tensor conservation equation
(\ref{conservation}). It is not difficult to see that the term~$X_p$, which arose from an integration by parts, can be removed by densitizing the equation; we may hence write
\begin{equation}\label{conserve}
\nabla^{LC}_p \tilde T_\textrm{eff}{}^p{}_i + |\textrm{Det }G|^{1/6}T_{abcd}\nabla^{LC}_i G^{abcd}=0\,.
\end{equation}
Note that the conservation of effective energy momentum
density depends on whether or not the second term in the equation
vanishes. In the almost metric case, this term
simplifies to
\begin{equation}\label{effcons}
\nabla^{LC}_p \tilde T_\textrm{eff}{}^p{}_i +
\sqrt{-g}\frac{\partial_i\phi}{\sqrt{1+\phi^2}}\,\omega_g^{abcd}T_{abcd}=0\,.
\end{equation}
Since $d\phi$ is generically non-zero, the effective energy
momentum density is only conserved if the matter source has no
totally antisymmetric component, i.e., if $\bar
T=\omega_g^{abcd}T_{abcd}=0$. As we will show below, this is the
case for an early universe filled with fermionic and bosonic
radiation.

The most relevant insight for our further developments in this article is that the physical momentum of the matter fields is given by
\begin{equation}\label{physmomentum}
  j^p = T_\textrm{eff}{}^p{}_0
\end{equation}
choosing a $g_G$-orthonormal frame $\{e_0,e_\alpha\}$ for an
observer with four-velocity $e_0 = u$. Note that for the
identification of radiation fields, whose physical momentum $j$ is
Fresnel-null, the normalization of the observer's velocity $u$ is
irrelevant.

Before continuing this discussion, we remark that $T_\textrm{eff}$
also seems to be meaningful as the physical energy-momentum of fields on
general area metric backgrounds, not only the almost metric ones
which are of direct relevance for our application to cosmology. An
explicit calculation on general area metric manifolds demonstrates
this for electrodynamics (\ref{nonabelian}): we will show
the equality of~$-\tilde T_\textrm{eff}{}^0{}_0$ to the
Hamiltonian energy density of the abelian gauge field $A$.
The calculation of the effective energy momentum tensor from the
source tensor (\ref{nonabeliansource}) yields
\begin{equation}
   \tilde T_\textrm{eff}{}^0{}_0 = |\textrm{Det }G|^{1/6}\Big(\frac{1}{2}G^{0\beta 0\delta}F_{0\beta}F_{0\delta} - \frac{1}{8}G^{\alpha\beta\gamma\delta}F_{\alpha\beta}F_{\gamma\delta}\Big),    
\end{equation}
denoting spatial indices by greek letters. The Hamiltonian density on the other hand is most easily obtained from the Lagrangian density $\mathcal{L}_A$ in the temporal gauge $A_0=0$, in which it becomes
\begin{equation}
   \mathcal{L}_A = -\frac{1}{8}|\textrm{Det }G|^{1/6}\left(G^{\alpha\beta\gamma\delta}F_{\alpha\beta}F_{\gamma\delta}+4G^{0\beta\gamma\delta}\dot A_\beta F_{\gamma\delta} +4G^{0\beta 0\delta}\dot A_\beta \dot A_\delta \right)\!.   
\end{equation}
The momentum conjugate to $A_\beta$ is defined as $\Pi^\beta=\partial\mathcal{L}_A/\partial \dot A_\beta$ and the Hamiltonian density is $\mathcal{H}=\Pi^\beta\dot A_\beta-\mathcal{L}_A$. One thus finds precise agreement, $\mathcal{H}=-\tilde T_\textrm{eff}{}^0{}_0$.

\section{More on null geometry}\label{sec_null}
Equipped with the physical momentum density for matter fields coupled to an area metric background, we would like to identify radiative solutions by null momentum. Recall from section~\ref{review} that null covectors are defined via the Fresnel tensor $\mathcal{G}^{ijkl}$, see (\ref{Fresneldef}), associated with the area metric. These Fresnel-null covectors have immediate physical significance as the directions of light fronts. For the identification of bosonic and fermionic radiation fields in the next section, however, we need to have at our disposal also a dual $\mathcal{G}_{ijkl}$ of the Fresnel tensor, in order to define radiation fields via null physical momentum vectors in the next section.

The basic idea leading to the construction of the dual Fresnel tensor may be borrowed
 directly from classical string theory; we know that the endpoints of open strings on metric manifolds follow null curves. We now carry this result over to open
strings on area metric manifolds: consider an open string
worldsheet $x(\tau,\sigma)$ with local tangent area $\Omega=\dot
x\wedge x'$ that solves the stationarity condition~\cite{Schuller:2005ru}
\begin{equation}
[dG^C(\Omega,\cdot)](\Omega,\cdot)=0\,,
\end{equation}
where $G^C$ denotes the cyclic part of the area metric, so that
this worldsheet becomes a minimal surface on $(M,G)$. We
parametrize the worldsheet so that one of its boundaries lies
at $\sigma=0$ and impose von~Neumann boundary conditions
${x'(\tau,0)=0}$. Since we are interested in the motion of the
boundary curve, we Taylor-expand around $\sigma=0$, which gives
\begin{equation}
x(\tau,\sigma)=y_0(\tau)+\frac{1}{2}\sigma^2 y_2(\tau) +\mathcal{O}(\sigma^3)\,,
\end{equation}
where we write $y_{i}(\tau)$ for the $i$-th derivative
$(\partial_\sigma)^i x(\tau,0)$. We substitute the
expansion into the stationarity condition. To lowest order
$\mathcal{O}(\sigma^0)$ this yields a single contribution, coming
from the second derivative~$x''$, which has to vanish on its own:
\begin{equation}
G^C_{abcd}(y_0)\dot y_0^a \dot y_0^c y_2^d=0\,.
\end{equation}
Choosing a basis $\{e_{\hat 0},e_{\hat \alpha}\}$ of the local
tangent spaces so that $e_{\hat 0}=\dot  y_0$, and using the
symmetries of~$G^C$, this equation becomes $G^C_{\hat
0\hat\beta\hat 0\hat\delta}y_2^{\hat\delta}=0$. Since $y_2\neq 0$
generically, the only way to satisfy this condition is to require
the vanishing of the determinant of $G^C_{\hat 0\hat\beta\hat
0\hat\delta}$, i.e., $\omega^{\hat
0\hat\alpha\hat\beta\hat\gamma}\omega^{\hat
0\hat\kappa\hat\lambda\hat\mu}G^C_{\hat 0\hat\alpha\hat
0\hat\kappa}G^C_{\hat 0\hat\beta\hat 0\hat\lambda}G^C_{\hat
0\hat\gamma\hat 0\hat\mu}=0$. Writing $X=\dot y_0$ we can obtain
this covariantly as
\begin{equation}\label{dualFresnel}
  \mathcal{G}_{abcd}X^aX^bX^cX^d=-\frac{1}{24}\omega_{G^C}^{ijkl}\omega_{G^C}^{mnpq}G^C_{ijm(a}G^C_{b|kn|c}G^C_{d)lpq}X^aX^bX^cX^d=0\,,
\end{equation}
which defines the totally symmetric covariant dual Fresnel tensor.
Note that our choice of the dual tensor $\omega_{G^C}$ implies
that the dual Fresnel tensor only depends on the algebraic
curvature part $G^C$ of the area metric. These results easily
generalize to higher dimensions, which however will not be needed
in the present paper.

We may thus call a vector $X$ on an area metric manifold null if it is null with respect to the dual Fresnel tensor according to equation (\ref{dualFresnel}). The same condition can be derived from the framework of pre-metric electrodynamics \cite{Peres:1962,Hehl:2004yk,Hehl:2005xu}, similarly as we did in \cite{Punzi:2006nx}, but starting from a geometric definition of light rays \cite{Kiehn:1991,Rubilar}. With the notion of null vectors on area metric manifolds at hand we finally turn to the discussion of radiation fields.

\section{Radiation fields}\label{sec_radiation}
Collecting results from the previous three sections, we may now provide an invariant characterization for radiation fields.  We start from the physical definition of a radiation field configuration as 
one for which the physical momentum (\ref{physmomentum}) is Fresnel-null everywhere,
\begin{equation}\label{radcond}
  \mathcal{G}_{abcd} j^a j^b j^c j^d = 0\,.
\end{equation}
For our application to cosmology, we are only interested in homogeneous and isotropic area metric manifolds, which take the almost metric form (\ref{almetric}), as discussed above; then the expression for the dual Fresnel tensor~(\ref{dualFresnel}) becomes
\begin{equation}
\mathcal{G}_{abcd}=\frac{g_{(ab}g_{cd)}}{1+\phi^2}\,,
\end{equation}
and the physical momentum conveniently simplifies to
\begin{equation}
j^p=8T^{cp}{}_{c0} + \omega_g^{ijkl}T_{ijkl}\,\phi\,\delta^p_0\,,
\end{equation}
compare (\ref{Teffsim}), where indices have been raised with the inverse metric $g^{-1}$, and where the macroscopic observer's frame defines the direction of time.
We will now discuss the specific cases of electrodynamics and Dirac spinors in turn.

For area metric electrodynamics, the source tensor is
given by (\ref{nonabelian}), with the trivial gauge group~$U(1)$.
The physical momentum thus becomes
\begin{equation}
j^p=F^{cp}F_{c0} + \frac{1}{2}\phi\,\omega_g(F,F)\delta^p_0 - \frac{1}{2}G^{-1}(F,F)\delta^p_0\,,
\end{equation}
where by our conventions $G^{-1}(F,F)=G^{abcd}F_{cd}F_{cd}/4$, and
similarly for the term with $\omega_g$. A rather lengthy
calculation, using the intermediate definitions
$F_{0\beta}=E_\beta$ for electric and
${F_{\alpha\beta}=\omega_{g\,0\gamma\alpha\beta}B^\gamma}$ for
magnetic components of the field strength, now reveals that
\begin{equation}
\mathcal{G}_{abcd}j^aj^bj^cj^d =
\frac{4g_{00}}{1+\phi^2}\left(C_{g^{-1}}(F,F)^2+\omega_g(F,F)^2\right)
\end{equation}
which must vanish for radiation. But since each of the field
invariants in the bracket is positive, they must vanish
separately. This in turn may be taken as an alternative definition
of radiation as solutions of area metric electrodynamics
characterized by vanishing field invariants $C_{g^{-1}}(F,F)=0$
and $\omega_g(F,F)=0$. Finally, we conclude that for gauge field
radiation, the totally antisymmetric contribution to the source
tensor vanishes,
\begin{equation}\label{omegaT}
\omega_G^{abcd}T_{abcd} = \frac{1}{2 (1+\phi^2)}
\,\omega_g(F,F)-\frac{\phi}{2 (1+\phi^2)}C_{g^{-1}}(F,F) = 0\,.
\end{equation}

For Dirac fermions we have the source tensor~(\ref{fermionsource}). Using the on-shell simplifications this source tensor implies an effective energy momentum tensor~(\ref{Teffsim}) of the form
\begin{equation}
T^\psi_\mathrm{eff}{}^a{}_b = \frac{i}{2}g^{ap}\bar\psi\gamma_{(p}D_{b)}\psi - \frac{i}{2}g^{ap}D_{(p}\bar\psi\gamma_{b)}\psi\,.
\end{equation}
Since $T_\mathrm{eff}{}^a{}_a=4 G^{abcd}T^\psi_{abcd}$, it immediately follows, again on-shell, that
\begin{equation}\label{psiGtrace}
  G^{abcd}T^\psi_{abcd} = \frac{i}{4} m\bar\psi\psi\,.
\end{equation}

Collecting the results (\ref{gaugetrace}), (\ref{omegaT}) for gauge fields, and
(\ref{psiGtrace}), (\ref{psiOmtrace}) for fermions, we thus arrive
at the conclusion that gauge field radiation and massless fermions on an
almost metric background satisfy the conditions
\begin{equation}\label{radcond_T}
   \omega_G^{abcd} T_{abcd} = 0\qquad \textrm{and} \qquad G^{abcd} T_{abcd} = 0\,.
\end{equation}
In the following section, we will derive the equations of state for a radiative string fluid from the vanishing of these two invariants.

\section{Radiation-dominated area metric cosmology}\label{sec_radcosmo}
Finally we turn to the effective string fluid that describes
radiation fields in area metric cosmology, both gauge fields and
ultrarelativistic fermions, which for all practical purposes
may be treated as massless. Imposing the radiation conditions
(\ref{radcond_T}) on the source tensor (\ref{stringfluid}) for a
general string fluid we obtain the equations of state for the
macroscopic variables $\tilde\rho$, $\tilde p$ and $\tilde q$ that
describe a radiation fluid:
\begin{equation}\label{radiation}
\tilde q = 0\,,\qquad \tilde\rho+\tilde p -2\tilde\rho \tilde\phi^2 = 0\,.
\end{equation}
These very simple equations of state present the technical key result of this paper.
Note here that the first relation $\tilde q=0$, equivalent to $\omega_G{}^{abcd} T_{abcd} = 0$, also guarantees the conservation of the effective energy momentum
(\ref{Tefftensor}), according to equation (\ref{effcons}).

We will now demonstrate that these relations for the macroscopic
string fluid variables imply that an area metric cosmology filled with bosonic and fermionic radiation evolves
 precisely as Einstein cosmology filled with a perfect radiation fluid.
As was shown in \cite{Punzi:2006nx}, the equations of motion for area metric cosmology (determined by a homogeneous and isotropic FLRW metric $g$ and scalar $\phi$) filled with a
general string fluid (with~$\tilde\rho$,~$\tilde p$ and~$\tilde q$) are precisely equivalent
 to the equations for Einstein cosmology filled with a perfect fluid
 whose energy density $\rho$ and pressure $p$ both depend in a rather intricate manner on the more fundamental variables of the string fluid,~$\tilde\rho$,~$\tilde p$ and~$\tilde q$, and on the scale factor $a$ and the scalar field~$\phi$. So,  schematically, we have
\begin{eqnarray} 
\textrm{Area cosmology } (g,\,\phi) & + & \textrm{String fluid } (\tilde\rho,\,\tilde p,\,\tilde q)\nonumber\\
& \Longleftrightarrow & \nonumber\\
\textrm{Einstein cosmology } (g) & + & \textrm{Perfect fluid } (\rho,\,p)\,. \nonumber
\end{eqnarray}
More precisely, one finds $\rho=3(x-y)$ and $p=x+y$, so that the effective equation of state parameter becomes
\begin{equation}\label{statepar}
w=\frac{p}{\rho}=\frac{x+y}{3(x-y)}\,,
\end{equation}
for
\begin{equation}\label{xy}
x=-H\dot{\tilde\phi}\tilde\phi^{-1}+4\kappa (\tilde\rho+\tilde
q)\tilde\phi^2\,,\quad y=4\kappa\tilde q\,,
\end{equation}
where $H=\dot a/a$ is the Hubble function and $\tilde\phi$ is
defined as in section \ref{sec_momentum}.
This mapping is a convenient formal trick that allows to compare the predictions
of area metric cosmology to standard cosmology. The
appearance of the gravitational degrees of freedom~$a$ and~$\tilde\phi$ in~(\ref{xy}), however, renders this map highly
non-trivial; in particular, it is the exception rather than the
rule that specific, physically meaningful equations of state for
the string fluid variables~$\tilde p,~\tilde\rho$ and~$\tilde q$
will recover the physically corresponding equations of state for a
perfect fluid.

For non-interacting string dust, for instance, we have shown in
\cite{Punzi:2006nx} that the equations of state take the form
$\tilde p = 0$ and $\tilde q = - \tilde\rho$; however, they do not
imply $p=0$ for the effective pressure, which fact lies at the
heart of the existence of the accelerating solution for the late
universe in area metric cosmology.

Our equations of state (\ref{radiation}) for radiation string
fluids, in contrast, imply that the  effective perfect fluid
indeed satisfies the familiar equation of state for radiation,
\begin{equation}
  p = \frac{1}{3} \rho\,,
\end{equation}
as one easily verifies by insertion of (\ref{radiation}) into
(\ref{statepar}). This immediately implies the equivalence of
area metric cosmology filled with radiation string fluids to
Einstein cosmology filled with a perfect radiation fluid. Of
course, area metric cosmology provides a more detailed solution
for the scalar field $\phi$ and the string fluid variables, but by
a miraculous cancellation these details do not in any way
affect the evolution of the scale factor.

Therefore the area metric cosmology of the early,
radiation-dominated universe is completely unchanged
with respect to Einstein cosmology, so that all successes,
such as for instance nucleosynthesis (yielding the correct
abundances of light elements already in standard cosmology), are
inherited. But as we saw in~\cite{Punzi:2006nx}, the late universe
in area metric cosmology does depart from Einstein cosmology in
allowing for the experimentally observed accelerating expansion.

\section{Conclusions}\label{conclusions}
Nowadays, cosmology provides an  excellent probe for our
theories of nature; first, due to the availability of
reliable, and partly unexpected, observational precision data, and second,
because of the intricate interplay between different branches of
fundamental physics that is needed to draw realistic conclusions.
While the non-trivial combination of general relativity and the
standard model of particle physics provides predictions consistent
with most data, the observed small acceleration of the late
universe indicates that our ideas about particle physics,
or gravity, or both, might have to be changed in order to
satisfactorily explain this stunning observation \cite{Peebles:2002gy,Copeland:2006wr,Nojiri:2006ri,Dvali:2003rk,Sahni:2002dx,Melchiorri:2002ux}.

The absence of a natural explanation within the cosmological
standard model suggests that there is something essential about
the interaction of spacetime and matter that we do not understand.
This lack of understanding may quickly be parametrized in form of
a cosmological constant, or a more elaborate model of dark energy;
but the inherent difficulty of such purely phenomenological
approaches is that we do not learn much at a fundamental level
from the failure or even success of any such particular model.

This insight is what has fuelled, at least from the point of view
of relativists, the excitement about string theory ever since it
became clear that both gravity and matter could arise from one
fundamental principle, and their interaction be determined.
Unfortunately, the consistency of quantum string theory only in
higher dimensions stands, to the present day, in the way of unique
phenomenological predictions. One qualitative aspect of quantum
string theory, however, which attracts increasing attention today,
are the refined effective geometries arising in form of two
additional massless fields, the Neveu-Schwarz two-form and the
dilaton, besides the graviton.

Area metric geometry is the geometrization of this insight: the
generalized effective backgrounds for strings can be neatly
described by an area metric. This restriction to a description of
only the massless modes, i.e., the refined effective background
geometry, comes with a great advantage. As we showed before, this
structure can be given consistent dynamics in four dimensions,
which we interpret as a refined gravity theory.

Intriguingly, the cosmology of this area metric gravity effortlessly provides a universe with increasingly small late-time acceleration, while the early, radiation-dominated epoch, as shown from first principles in this paper, evolves like in standard cosmology. Both the early and late time behaviour are thus consistent with observations, and follow from the single principle of an area metric spacetime structure. Especially since the derivation of these predictions also required an understanding of the consistent coupling of bosonic and fermionic matter, they provide 
non-trivial support for the consistency of an area metric structure of physical spacetime.

More precisely, in this paper we studied the consistent coupling of Dirac spinors and non-abelian gauge fields to an area metric. The deeper insights into the null geometry of area metric manifolds, afforded here by studying open strings, were instrumental in providing a physically meaningful definition of radiation on area metric spacetimes. We showed that both radiation gauge fields and massless Dirac fermions are characterized by the vanishing of two invariants of their on-shell source tensors, which refine the notion of an energy-momentum tensor in the context of area metric spacetime. The technical key result of the paper, namely the equations of state for a string fluid describing bosonic and fermionic radiation, was then derived by imposing that the same invariants vanish for the source tensor of a string fluid. Other than in the case of a string dust fluid, which leads to a late-time acceleration of the universe, the string radiation fluid does not induce any deviation from Einstein cosmology in the early, radiation-dominated epoch of the universe. This shows that the success of the theory in explaining cosmic acceleration does not come at the cost of
inconsistencies in the early universe.     

Maybe the most desirable feature of the theory is its direct falsifiability, since no undetermined new fundamental constant is introduced. The falsifiability is a direct merit of using a refined geometry; on a metric manifold, in contrast, any modification of the standard gravitational action requires the introduction of a length scale for dimensional reasons alone. Pushing this length scale to smaller and smaller values 
may then always be used to achieve compatibility with standard predictions within any experimental margin of error.

In contrast, every single prediction made by area metric gravity provides a rigid check on its validity. The next challenge is to derive predictions for the solar system, where the reduced symmetry allows for even more deviation from purely metric backgrounds, as compared to cosmology. 
One immediate consequence is the kinematical possibility for arbitrarily large bi-refringence, which however is tightly constrained by readily available high precision data. It has to be seen whether the gravitational field equations constrain the solutions in such a way that bi-refringence is either absent or sufficiently highly suppressed. Again, the absence of a freely adjustable scale makes this another really decisive test of the area metric hypothesis.    

\acknowledgments
The authors wish to thank Alexander Turbiner, Daniel Sudarsky, Raffaele Punzi and Marcus Werner for helpful discussions. MNRW thanks the Instituto de Ciencias Nucleares, where this work was begun, for their warm hospitality, and acknowledges full financial support through the Emmy Noether Fellowship grant WO 1447/1-1 from the German Research Foundation DFG. 

\appendix
\section{Conventions}\label{app_Dirac}
In contrast to most treatments of quantum field theory we employ a mainly plus signature convention for the Lorentzian metric; our metric Riemann tensor is defined, in components, as $R^a{}_{bcd} = 2\Gamma^a_{b[d,c]} + 2\Gamma^a_{p[c}\Gamma^p_{|b|d]}$, and the Ricci tensor as $R_{bd}=R^a{}_{bad}$. Lagrangian densities in the matter action $S_m=\int \mathcal{L}_m$ are written with positive kinetic terms; then the Hamiltonian $\mathcal{H}=-T^0{}_0$, for the energy momentum tensor $T^{ab}=2/\sqrt{-g}\,\delta S_m/\delta g_{ab}$. Also, $T_{00}>0$, so that the Einstein equations take the form $R_{ab}-Rg_{ab}/2=8\pi G_NT_{ab}$.

For specific calculations with our conventions, we now also list consistent choices for the treatment of spinors. Together with the two-dimensional identity matrix $\openone_2$, the Pauli sigma matrices
\begin{equation}
\sigma^1=\Bigg(\begin{array}{cc}0&1\\1&0\end{array}\Bigg),\qquad\sigma^2=\Bigg(\begin{array}{cc}0&-i\\i&0\end{array}\Bigg),\qquad\sigma^3=\Bigg(\begin{array}{cc}1&0\\0&-1\end{array}\Bigg)
\end{equation}
form a basis of the vector space of two-dimensional complex Hermitian matrices. They satisfy the anticommutator relation $\{\sigma^\alpha,\sigma^\beta\}=2\delta^{\alpha\beta}\openone_2$. Using the notation
\begin{equation}
\sigma^\mu=(\openone_2,\sigma^\alpha)\,,\qquad \bar\sigma^\mu=(-\openone_2,\sigma^\alpha)\,,
\end{equation}
the Pauli matrices furnish us with a representation of the Dirac algebra on flat spacetime:
\begin{equation}
\Gamma^\mu=\Bigg(\begin{array}{cc}0&\sigma^\mu\\ \bar\sigma^\mu&0\end{array}\Bigg).
\end{equation}
Indeed, the anticommutator relation of the Pauli matrices immediately translates into the well-known anticommutator of Dirac gamma matrices,
\begin{equation}\label{Diracalg}
\{\Gamma^\mu,\Gamma^\nu\}=2\eta^{\mu\nu}\openone_4\,,
\end{equation}
for flat spacetime metric $\eta$ with mainly plus signature $(1,3)$. This signature is a consequence of the definition of $\bar\sigma^\mu$ which differs by a sign from that in most quantum field theory texts. Note that this representation is defined up to unitary equivalence; any redefined set of matrices $\Gamma'{}^\mu=U\Gamma^\mu U^\dagger$ for unitary $U$ with $U^\dagger U=\openone_4$ satisfies the same Clifford algebra.  It is very useful to have a simple expression for the Hermitian conjugate of the Dirac matrices. In the representation introduced above one easily checks that
\begin{equation}\label{Hermi}
\Gamma^\mu{}^\dagger=\Gamma^0\Gamma^\mu\Gamma^0\,,
\end{equation}
but this statement is independent of an arbitrary unitary change of representation.

The commutator of the Dirac matrices gives, as usual, the generators of the Lorentz algebra in the spinor representation:
\begin{equation}\label{Diracgen}
\Sigma^{\mu\nu}=\frac{1}{4} [\Gamma^\mu,\Gamma^\nu]\,,
\qquad
{}[\Sigma^{\mu\nu},\Sigma^{\rho\sigma}] = \eta^{\rho\mu}\Sigma^{\sigma\nu} - \eta^{\rho\nu}\Sigma^{\sigma\mu}-(\rho\leftrightarrow\sigma)\,.
\end{equation}
A representation of the proper orthochronous Lorentz group, which is the subgroup continuously connected to the identity, is then given by
\begin{equation}\label{grouprep}
\Lambda_{1/2}(\omega)=\exp \Big(\frac{1}{2}\omega_{\mu\nu}\Sigma^{\mu\nu}\Big)
\end{equation}
for real antisymmetric parameters $\omega_{\mu\nu}$. Note that $\Lambda_{1/2}(\omega){}^{-1}=\Lambda_{1/2}(-\omega)$; using this fact it can be shown in standard fashion that the Dirac matrices transform as a Lorentz vector:
\begin{equation}\label{vectra}
\Lambda_{1/2}{}^{-1}\Gamma^\mu\Lambda_{1/2}=\Lambda^\mu{}_\rho\Gamma^\rho\,.
\end{equation}
From the simple behaviour of the gamma matrices under Hermitian conjugation, it follows that $\Sigma^{\mu\nu}{}^\dagger=\Gamma^0\Sigma^{\mu\nu}\Gamma^0$, and hence that $\Lambda_{1/2}{}^\dagger=-\Gamma^0 \Lambda_{1/2}{}^{-1}\Gamma^0$. The Lorentz transformation of Dirac spinors $\psi$ and $\bar\psi=\psi^\dagger\Gamma^0$ then becomes
\begin{equation}
\psi \mapsto \Lambda_{1/2}\psi\,,\qquad
\bar\psi \mapsto \bar\psi \Lambda_{1/2}{}^{-1}\,.
\end{equation}
This allows us to write down the Lagrangian density for massive Dirac spinors on flat space:
\begin{equation}\label{Diracact}
\mathcal{L}_\psi = \frac{1}{2}i\bar\psi \Gamma^\mu\partial_\mu \psi - \frac{1}{2}i \partial_\mu \bar\psi \Gamma^\mu \psi - i m \bar\psi \psi\,.
\end{equation}
The form of $\mathcal{L}$ is chosen symmetric in $\psi$ and $\bar\psi$ for convenience of generalization in the paper; the odd factors of $i$ ensure that $\mathcal{L}$ is Hermitian, and thus real.

\thebibliography{00}
\bibitem{Punzi:2006hy}
  R.~Punzi, F.~P.~Schuller and M.~N.~R.~Wohlfarth,
  arXiv:hep-th/0612133.

\bibitem{Punzi:2006nx}
  R.~Punzi, F.~P.~Schuller and M.~N.~R.~Wohlfarth,
  JHEP {\bf 0702}, 030 (2007)
  [arXiv:hep-th/0612141].
  
\bibitem{Sarkar:1995dd}
  S.~Sarkar,
  Rept.\ Prog.\ Phys.\  {\bf 59} (1996) 1493
  [arXiv:hep-ph/9602260].

\bibitem{Tytler:2000qf}
  D.~Tytler, J.~M.~O'Meara, N.~Suzuki and D.~Lubin,
  Phys.\ Scripta {\bf T85} (2000) 12
  [arXiv:astro-ph/0001318].

\bibitem{Spergel:2003cb}
  D.~N.~Spergel {\it et al.}  [WMAP Collaboration],
  Astrophys.\ J.\ Suppl.\  {\bf 148} (2003) 175
  [arXiv:astro-ph/0302209].

\bibitem{Knop:2003iy}
  R.~A.~Knop {\it et al.}  [Supernova Cosmology Project Collaboration],
  Astrophys.\ J.\  {\bf 598} (2003) 102
  [arXiv:astro-ph/0309368].


\bibitem{Hitchin:2004ut}
  N.~Hitchin,
  Quart.\ J.\ Math.\ Oxford Ser.\  {\bf 54} (2003) 281
  [arXiv:math.dg/0209099].

\bibitem{Gualtieri:2004}
  M.~Gualtieri,
  {\sl Generalized complex geometry},
  Oxford University DPhil thesis,
  [arXiv:math.dg/0401221].

\bibitem{Flournoy:2004vn}
  A.~Flournoy, B.~Wecht and B.~Williams,
  Nucl.\ Phys.\ B {\bf 706} (2005) 127
  [arXiv:hep-th/0404217].

\bibitem{Hull:2004in}
  C.~M.~Hull,
  JHEP {\bf 0510} (2005) 065
  [arXiv:hep-th/0406102].

\bibitem{Dabholkar:2005ve}
  A.~Dabholkar and C.~Hull,
  JHEP {\bf 0605} (2006) 009
  [arXiv:hep-th/0512005].

\bibitem{Grana:2004bg}
  M.~Grana, R.~Minasian, M.~Petrini and A.~Tomasiello,
  JHEP {\bf 0408} (2004) 046
  [arXiv:hep-th/0406137].

\bibitem{Grana:2005ny}
  M.~Grana, J.~Louis and D.~Waldram,
  JHEP {\bf 0601} (2006) 008
  [arXiv:hep-th/0505264].

\bibitem{Koerber:2005qi}
  P.~Koerber,
  JHEP {\bf 0508} (2005) 099
  [arXiv:hep-th/0506154].

\bibitem{Zucchini:2005rh}
  R.~Zucchini,
  JHEP {\bf 0503} (2005) 022
  [arXiv:hep-th/0501062].

\bibitem{Zabzine:2006uz}
  M.~Zabzine,
  arXiv:hep-th/0605148.

\bibitem{Reid-Edwards:2006vu}
  R.~A.~Reid-Edwards,
  arXiv:hep-th/0610263.

\bibitem{Grange:2006es}
  P.~Grange and S.~Schafer-Nameki,
   Nucl.\ Phys.\  B {\bf 770}, 123 (2007)
  [arXiv:hep-th/0609084].

\bibitem{Becker:2006ks}
  K.~Becker, M.~Becker, C.~Vafa and J.~Walcher,
  Nucl.\ Phys.\  B {\bf 770}, 1 (2007)
  [arXiv:hep-th/0611001].

\bibitem{Schuller:2005ru}
  F.~P.~Schuller and M.~N.~R.~Wohlfarth,
  JHEP {\bf 0602}, 059 (2006)
  [arXiv:hep-th/0511157].

\bibitem{Peres:1962}
  A.~Peres,
  Ann.\ Phys.\ {\bf 19} (1962) 279.

\bibitem{Hehl:2004yk}
  F.~W.~Hehl and Y.~N.~Obukhov,
  Found.\ Phys.\  {\bf 35} (2005) 2007
  [arXiv:physics/0404101].

\bibitem{Hehl:2005xu}
  F.~W.~Hehl, Yu.~N.~Obukhov, G.~F.~Rubilar and M.~Blagojevic,
  Phys.\ Lett.\ A {\bf 347} (2005) 14
  [arXiv:gr-qc/0506042].
  
\bibitem{Krasnov:2007uu}
  K.~Krasnov,
  arXiv:gr-qc/0703002.
  
\bibitem{Krasnov:2007ky}
  K.~Krasnov and Y.~Shtanov,
  arXiv:0705.2047 [gr-qc].

\bibitem{Cartan:1933}
  E.~Cartan,
  {\sl Les espaces m\'etriques fond\'es sur la notion d'aire},
  Hermann, Paris 1933.
  
\bibitem{Kiehn:1991}
  R.~M.~Kiehn, G.~P.~Kiehn and J.~B.~Roberds,
  Phys.\ Rev.\ A {\bf 43} (1991) 5665.
  
\bibitem{Rubilar}
G. Rubilar, {\sl Linear pre-metric 
electrodynamics and deduction of the lightcone}, Cologne University PhD thesis, Ann.\ Phys.\ (Leipzig) {\bf 11} (2002) [10-11] 717-782. 

\bibitem{Peebles:2002gy}
  P.~J.~E.~Peebles and B.~Ratra,
  Rev.\ Mod.\ Phys.\  {\bf 75} (2003) 559
  [arXiv:astro-ph/0207347].

\bibitem{Copeland:2006wr}
  E.~J.~Copeland, M.~Sami and S.~Tsujikawa,
  Int.\ J.\ Mod.\ Phys.\  D {\bf 15} (2006) 1753
  [arXiv:hep-th/0603057].

\bibitem{Nojiri:2006ri}
  S.~Nojiri and S.~D.~Odintsov,
  Int.\ J.\ Geom.\ Meth.\ Mod.\ Phys.\  {\bf 4} (2007) 115
  [arXiv:hep-th/0601213].

\bibitem{Dvali:2003rk}
  G.~Dvali and M.~S.~Turner,
  arXiv:astro-ph/0301510.

\bibitem{Sahni:2002dx}
  V.~Sahni and Y.~Shtanov,
  JCAP {\bf 0311} (2003) 014
  [arXiv:astro-ph/0202346].

\bibitem{Melchiorri:2002ux}
  A.~Melchiorri, L.~Mersini-Houghton, C.~J.~Odman and M.~Trodden,
  Phys.\ Rev.\  D {\bf 68} (2003) 043509
  [arXiv:astro-ph/0211522].
\end{document}